\documentclass{article}
\usepackage{icmc2026_paper_template}
\usepackage{times}
\usepackage{ifpdf}
\usepackage{soul}
\usepackage[english]{babel}
\usepackage{booktabs}
\usepackage{amsmath}
\usepackage{multirow, makecell}
\usepackage{microtype}
\usepackage{xcolor}
\usepackage{float}
\usepackage{mathtools}
\usepackage{enumitem}
\usepackage{amssymb}
\usepackage[table]{xcolor}
\def\papertitle{Snapping Matters: Context-Aware Onset Refinement for Automatic Music Transcription}

\def\firstauthor{Abhirup Saha}
\def\secondauthor{Hans-Ulrich Berendes}
\def\thirdauthor{Meinard Müller}
\def\fourthauthor{Ben Maman}
\newif\ifpdf
\ifx\pdfoutput\relax
\else
   \ifcase\pdfoutput
      \pdffalse
   \else
      \pdftrue
  \fi
\fi

\ifpdf % compiling with pdflatex
  \usepackage[pdftex,
    pdftitle={\papertitle},
    pdfauthor={\firstauthor, \secondauthor, \thirdauthor},
    bookmarksnumbered, % use section numbers with bookmarks
    pdfstartview=XYZ % start with zoom=100% instead of full screen; 
   ]{hyperref}
  \usepackage[pdftex]{graphicx}
  \graphicspath{{./figures/}}
  \DeclareGraphicsExtensions{.pdf,.jpeg,.png}

  \usepackage[figure,table]{hypcap}

\else % compiling with latex
  \usepackage[dvips,
    bookmarksnumbered, % use section numbers with bookmarks
    pdfstartview=XYZ % start with zoom=100% instead of full screen
  ]{hyperref}  % hyperrefs are active in the pdf file after conversion

  \usepackage[dvips]{epsfig,graphicx}
  \graphicspath{{./figures/}}
  \DeclareGraphicsExtensions{.eps}

  \usepackage[figure,table]{hypcap}
\fi

\hypersetup{
    colorlinks,%
    citecolor=black,%
    filecolor=black,%
    linkcolor=black,%
    urlcolor=black
}

\title{\papertitle}

\fourauthors
   {0.1in}
   {\firstauthor} {%
     \parbox[c]{3in}{\centering
       International Audio Laboratories Erlangen\\
       {\tt \href{mailto:abhirup.saha@audiolabs-erlangen.de}{\small abhirup.saha@audiolabs-erlangen.de}}
     }%
   }
   {\secondauthor} {%
     \parbox[c]{3in}{\centering
       International Audio Laboratories Erlangen\\
       {\tt \href{mailto:hans-ulrich.berendes@audiolabs-erlangen.de}{\small hans-ulrich.berendes@audiolabs-erlangen.de}}
     }%
   }
   {\thirdauthor} {%
     \parbox[c]{3in}{\centering
       International Audio Laboratories Erlangen\\
       {\tt \href{mailto:meinard.mueller@audiolabs-erlangen.de}{\small meinard.mueller@audiolabs-erlangen.de}}
     }%
   }
   {\fourthauthor} {%
     \parbox[c]{3in}{\centering
       International Audio Laboratories Erlangen\\
       {\tt \href{mailto:ben.maman@audiolabs-erlangen.de}{\small ben.maman@audiolabs-erlangen.de}}
     }%
   }

\begin{document}
\capstartfalse
\maketitle
\capstarttrue

\begin{abstract}
Precise note-level annotations are critical for training automatic music transcription (AMT) systems, in particular note-onset labels, which form a core component of many recent AMT systems. However, high-quality annotations for real-world recordings are scarce. Sequence-level score--audio alignment methods such as dynamic time warping provide only coarse correspondence, making a local refinement step necessary. This refinement step, known as snapping, adjusts aligned score onsets using peaks in a neural onset posteriorgram and often determines whether weakly aligned score--audio pairs become usable training data at all. Despite its practical importance, snapping is typically treated as a simple post-processing heuristic and implemented with greedy local decisions. We present a systematic analysis of snapping strategies for training instrument-agnostic transcribers, demonstrating that snapping is essential for learning from weakly aligned data. Building on this, we formulate snapping as a per-pitch assignment problem and solve it via bipartite graph matching, yielding context-aware onset decisions under overlapping refinement windows and uncertain initial alignments. Extensive cross-dataset experiments across piano, chamber, and orchestral recordings show improved onset alignment and transcription accuracy over greedy snapping, with gains increasing for wider snapping windows and coarser initial alignments. Qualitative examples are provided on our project page: \url{https://abhirupsaha8.github.io}

\end{abstract}

\section{Introduction}\label{sec:introduction}

\begin{figure}
    \centering
    \includegraphics[width=\linewidth]{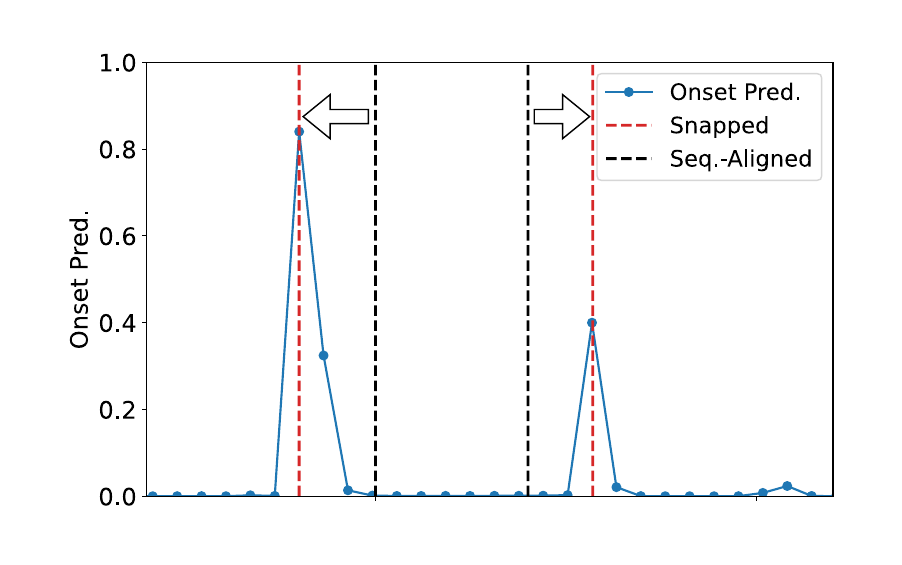}
    \vspace{-1cm}
    \caption{
    Snapping for a single pitch. The black dashed line (``Seq.-Aligned'') shows onset timings from a sequence-level alignment (e.g., DTW), while the red dashed line (``Snapped'') indicates the adjusted onsets after snapping. The blue line denotes
    activations in an onset posteriorgram.
    % onset activations in the posteriorgram that guide the snapping process.
}
    \label{fig:teaser}
\end{figure}

Automatic music transcription (AMT), the task of converting audio recordings into symbolic note representations, remains a central problem in music information retrieval (MIR). While piano transcription has progressed rapidly, reliable onset and pitch estimation for non-piano instruments and multi-instrument mixtures is still difficult due to timbral variability, polyphony, and the limited availability of precisely annotated training data. In particular, modern data-driven AMT models rely on supervision with accurate note onsets, yet onset-level ground truth is hard to obtain for real recordings outside controlled acquisition settings.

A common workaround is to derive training labels by aligning a musical score (or a MIDI file) to the audio. Sequence-level alignment methods such as dynamic time warping (DTW) can provide robust coarse correspondences, for example based on chroma- or onset-related features~\cite{EwertMG09_HighResAudioSync_ICASSP, MuellerOKPD21_SyncToolbox_JOSS, ZeitlerMM24_Synchronization_ISMIR}. However, even strong sequence-level alignments typically do not yield onset-accurate labels. Expressive timing deviations and local asynchronies (e.g., arpeggiation) can cause score events that are simultaneous in the score to be realized at different times in the audio, or vice versa. Consequently, directly transferring score onsets through a warping path often yields labels that are insufficiently precise for effective AMT training.

To bridge this gap, many recent pipelines adopt a two-stage strategy~\cite{HuD06_bootstrapOnset_ML, MamanBermano22_UnalignedAMT_ICML, RileyGED2024_GAPS_ISMIR}: first compute a global, sequence-level score--audio alignment, then refine individual note onsets using neural onset activations, a step commonly referred to as snapping. In snapping, each score onset timing is refined within a local temporal window to a nearby peak in a learned onset posteriorgram, as depicted in Figure~\ref{fig:teaser}. Despite its practical success and growing adoption, snapping is typically implemented using local greedy decisions or simplified peak-picking heuristics. This is fragile when refinement windows overlap, because locally optimal choices can conflict, leading to duplicate assignments, missed peaks, or reduced global consistency. The problem becomes particularly pronounced when the initial alignment is coarse and wider refinement windows are required.

This paper revisits \emph{snapping} as an essential component of AMT label generation and asks how onset refinement should be formulated and solved to remain reliable under realistic alignment uncertainty. Our key idea is to treat snapping not as independent peak picking, but as a structured decision problem. For each pitch, we assign note events to candidate audio frames within admissible time windows, maximizing onset-posterior evidence while enforcing one-to-one consistency. This yields a principled and scalable refinement step that remains robust in the presence of overlapping windows and coarse initial alignments. The main contributions of this paper are as follows:
\begin{itemize}[leftmargin=*]
\item We clearly distinguish \emph{sequence-level alignment}, which defines a warping relation between timelines, from \emph{note-onset-level alignment}, which maps discrete events to precise times, and we position snapping as the refinement step that connects the two.
\item We formulate snapping as a per-pitch assignment problem based on bipartite graph matching, explicitly handling overlapping windows and global consistency beyond greedy heuristics.
\item We provide a systematic cross-dataset evaluation analyzing how solver choice, window size, and initial alignment quality affect both alignment accuracy and downstream transcription performance, including robustness under very coarse initial alignments.
\item We show that graph-based snapping yields consistent improvements in transcription accuracy across diverse datasets and becomes increasingly beneficial as refinement windows grow, which is precisely the regime required when labels are only weakly aligned.
\end{itemize}

This establishes snapping as a principled and tunable refinement step that bridges robust sequence alignment and onset-accurate supervision, enabling more reliable training for instrument-agnostic AMT under realistic conditions.

The remainder of this paper is organized as follows. Section 2 reviews related work on AMT and audio–score alignment. Section 3 introduces our method, including formal definitions of sequence- and onset-level alignment and a mathematical formulation of graph-based snapping. Section 4 defines the transcription and alignment tasks considered in this work. Section 5 presents experiments and evaluation procedures. Section 6 summarizes the findings and outlines directions for future work.

\begin{figure*}[t!]
    \centering
    \scalebox{2}{
        \includegraphics[width=\columnwidth]{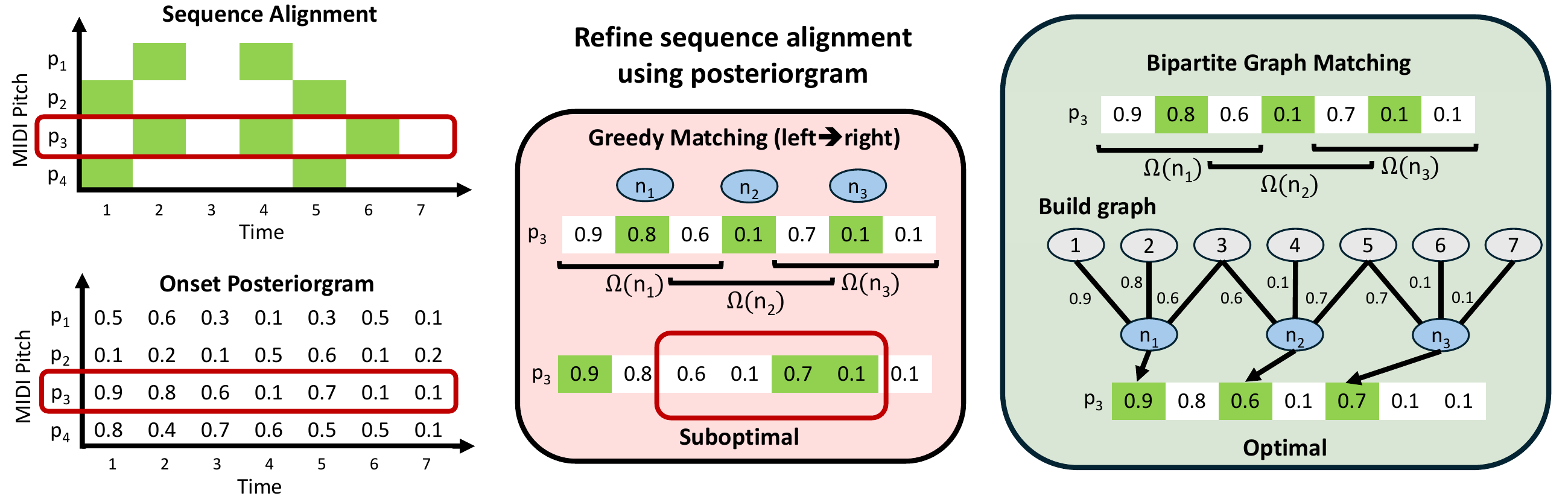}
    }
    \caption{
    Snapping using greedy matching, compared to snapping based on bipartite graph matching.
    }

    \label{fig:overview}
\end{figure*}

\section{Related Work}\label{sec:related}
For an overview of automatic music transcription, see~\cite{BenetosDDE19_MusicTranscription_SPM}. Early work focused on piano transcription, initially using non-negative matrix factorization (NMF)~\cite{SmaragdisB03_MusicTranscriptionNMF_WASPAA}, before giving way to data-driven approaches. Key advances such as Onsets and Frames~\cite{HawthorneESRSRE18_OnsetsFrames_ISMIR} and the MAESTRO dataset~\cite{HawthorneSRSHDE19_MAESTRO_ICLR} enabled high-quality polyphonic piano transcription~\cite{KongEtAl21_HighResTranscription_TASLP} based on note-onset detection.

For multi-instrument datasets~\cite{ThickstunHK17_MusicNet_ICLR}, score--audio alignment via DTW~\cite{MuellerOKPD21_SyncToolbox_JOSS} is common but prone to errors. Recent work~\cite{MamanBermano22_UnalignedAMT_ICML} shows that onset features from pre-trained neural transcription models enable more precise score--audio alignment, improving sequence-level robustness~\cite{ZeitlerMM24_Synchronization_ISMIR} and allowing accurate note-onset annotations via snapping. A similar strategy was later used to construct guitar datasets~\cite{RileyED24_GuitarTranscription_ICASSP, RileyGED2024_GAPS_ISMIR}. More recent work further simplifies alignment using histogram-based top-$K$ peak picking~\cite{YaffeMMB25_CountEM_ISMIR}.

While initially applied to piano and other instruments with sharp onsets such as guitar, onset-based transcription has been extended in recent work to strings, winds, and multi-instrument settings, both instrument-sensitive~\cite{WuCS20_TranscriptionSelfattn_TASLP, GardnerSMHE22_MultiTaskTranscription_ICLR, MamanBermano22_UnalignedAMT_ICML} and instrument-agnostic~\cite{WuWLLYGL24_Harmonic_ICME, YaffeMMB25_CountEM_ISMIR}.

\section{Method}\label{sec:method}
In this section, we describe our proposed method in detail. In Section~\ref{sec:seq_vs_note}, we formally define the notions of sequence-level and note-onset-level alignment, highlighting the differences between them. Then, in Section~\ref{sec:snapping}, we formulate how snapping---the refinement of a sequence-level alignment into an onset-level alignment---can be performed optimally using bipartite graph matching based on a learned note-onset posteriorgram.
\subsection{Sequence-level vs. Note-Onset-level Alignment}\label{sec:seq_vs_note}
Given an audio recording of a musical piece and a corresponding digital representation of the score, such as a piano roll or a MIDI file, we define two discrete timelines: the score timeline
\begin{equation}
\mathcal{T}_{\mathrm{s}} = [1: T_\mathrm{s}]\coloneqq \{1,2,\dots,T_\mathrm{s}\}
\end{equation}
of length \(T_\mathrm{s}\in\mathbb{N}\), and the audio (physical) timeline
\begin{equation}
\mathcal{T}_{\mathrm{a}} = [1: T_\mathrm{a}]\coloneqq \{1,2,\dots,T_\mathrm{a}\}
\end{equation}
% In this work, we adopt a temporal resolution of 31.25 frames per second.
of length \(T_\mathrm{a}\in\mathbb{N}\).
We further define a set of possible pitches \(\mathcal{P}\) of size $P$, and a set of possible instruments \(\mathcal{I}\) of size $I$:
\begin{equation}
\mathcal{P} = \{p_1,\dots,p_P\},\quad \mathcal{I} = \{i_1,\dots i_I\}.
\end{equation}
% Using these definitions,
The musical score can then be represented as a set of notes
\begin{equation}\label{eq:note_list}
\mathcal{N}\subset \mathcal{P}\times\mathcal{T}_{\mathrm{s}}\times \mathcal{T}_{\mathrm{s}}\times \mathcal{I}
\end{equation}
of size \(N\).
% \begin{equation}\label{eq:note_list}
% \mathcal{N} = \{n_1,\dots,n_N\}
% % ,\quad
% % n = (p_k,\, t_k^{\mathrm{on}},\, t_k^{\mathrm{off}},\, i_k).
% \end{equation}
Each note \(n\in\mathcal{N}\) is represented as a tuple
\begin{equation}\label{eq:note_format}
n = (p,\, t^{\mathrm{on}},\, t^{\mathrm{off}},\, i)
\end{equation}
where \(p\in\mathcal{P}\) denotes the pitch, \(t^{\mathrm{on}}\in \mathcal{T}_{\mathrm{s}}\) and \(t^{\mathrm{off}}\in \mathcal{T}_{\mathrm{s}}\) are the note onset and offset positions on the score timeline, and \(i\in\mathcal{I}\) is the instrument class.
% In this work we focus on instrument-agnostic transcription (see Section~\ref{sec:agnostic}) and therefore discard instrument information. 

A \emph{sequence-level alignment} is defined as a function
\begin{equation}
\mathcal{W}: \mathcal{T}_{\mathrm{s}} \to \mathcal{T}_{\mathrm{a}},
\end{equation}
which maps each score-frame index to a corresponding audio-frame index. Such alignments are typically estimated from frame-level similarity using DTW or similar methods.

% It is defined as a pair of functions
% \begin{equation}
% (\mathcal{M}^{\mathrm{on}}, \mathcal{M}^{\mathrm{off}}): \mathcal{N} \to \mathcal{T}_{\mathrm{a}} \times \mathcal{T}_{\mathrm{a}}.
% \end{equation}
% In this work we focus on onset detection, and therefore consider only the onset component. We define an onset-level alignment
% and offset positions
A \emph{note-onset-level alignment} is a function
\begin{equation}
\mathcal{M}: \mathcal{N} \to \mathcal{T}_{\mathrm{a}},
\end{equation}
mapping each note in the musical score \(\mathcal{N}\) to its corresponding onset time on the audio timeline \(\mathcal{T}_{\mathrm{a}}\). In this work, we focus on onset detection and consider only the onset component of each note, 
although a more general note-level alignment could also include note offsets.
% A note-onset-level alignment assigns to each note its onset position
% on the audio timeline. It is defined as a function
% \begin{equation}
% \mathcal{M}: \mathcal{N} \to \mathcal{T}_{\mathrm{a}},
% \end{equation}
% which maps each note to its onset time in the physical timeline. In this work we focus on onset detection, and therefore consider only the onset component of a note. However, in principle, a broader definition of note-level alignment could be used including also mapping each note to its offset time in the physical timeline.

A sequence-level alignment \(\mathcal{W}\) induces an onset-level alignment $\mathcal{M}^\mathrm{seq}:\mathcal{N}\to\mathcal{T}_\mathrm{a}$ defined by:
\begin{equation}
\mathcal{M}^\mathrm{seq}(n) = \mathcal{M}^\mathrm{seq}((p,\, t^{\mathrm{on}},\, t^{\mathrm{off}},\, i)) = \mathcal{W}(t^{\mathrm{on}}).
\end{equation}

When aligning a musical score to an audio recording of a performance of the same score, a good sequence-level alignment does not necessarily guarantee a good onset-level alignment. For example, if two note onsets occur within the same frame on the score timeline but in different frames on the audio timeline (as in arpeggios or similar expressive timing variations), then any onset-level alignment induced directly from the sequence-level alignment will necessarily assign an incorrect onset time to at least one of the notes.

% Nevertheless, if we denote
Denoting the true (latent) onset-level alignment by
\begin{equation}
\mathcal{M}^* : \mathcal{N} \to \mathcal{T}_\mathrm{a},
\end{equation}
% by \(\mathcal{M}^*\)
and the onset-level alignment induced by the sequence-level alignment by \(\mathcal{M}^\mathrm{seq}\), we assume in the following the existence of a uniform error bound \(B\) such that
\begin{equation}\label{eq:bound}
\lvert \mathcal{M}^\mathrm{seq}(n) - \mathcal{M}^*(n) \rvert \le B,
\qquad \text{for all } n \in \mathcal{N}.
\end{equation}
% \begin{equation}
% \lvert \mathcal{M}(n) - \mathcal{M}^*(n) \rvert \le B,
% \qquad \forall\, n \in \mathcal{N}.
% \end{equation}
Improved sequence-level alignment results in a smaller \(B\).

We refer to \emph{snapping} as the process of refining a sequence-level alignment into an onset-level alignment, i.e., estimating \(\mathcal{M}^*\) from \(\mathcal{M}^\mathrm{seq}\), by incorporating additional information from a note-onset posteriorgram.

% , injective (one-to-one)
% We assume that $\mathcal{M}^*_p$ is injective (one-to-one), i.e., if a note with pitch $p$ occurs in different times in the score it will occur in different times in the audio.
% $\mathcal{N}_p=\{n_1,\dots,n_{K_p}\}$
% We posit the existence of a latent, injective (one-to-one) true onset-level alignment, or \emph{matching} for the pitch $p$:
% \begin{equation}
% \mathcal{M}_p^* : \mathcal{N}_p \to \mathcal{T}_\mathrm{a}
% \end{equation}
% which we aim to estimate.
% Without loss of generality, we assume $\mathcal{M}^\mathrm{seq}$ is monotonic, i.e., $k<j\implies \mathcal{M}^\mathrm{seq}(n)< \mathcal{M}^\mathrm{seq}(n_j)$.
\subsection{Snapping as Bipartite Graph Matching}\label{sec:snapping}
We formulate \emph{snapping} as a \emph{bipartite graph-matching} problem, applied independently to each pitch, as illustrated in Figure~\ref{fig:overview} (right). For each pitch \( p \in \mathcal{P} \), let \( \mathcal{N}_p \subseteq \mathcal{N} \) denote the subset of notes whose pitch is \( p \). We define
\begin{equation}
\mathcal{M}_p^* = \mathcal{M}^*|_{\mathcal{N}_p} : \mathcal{N}_p \to \mathcal{T}_\mathrm{a}.
\end{equation}
We estimate \( \mathcal{M}_p^* \) with an onset-level alignment
\begin{equation}
\mathcal{M}_p : \mathcal{N}_p \to \mathcal{T}_\mathrm{a},
\end{equation}
and subsequently unify these pitch-wise mappings to obtain a single onset-level alignment over the full note set,
\begin{equation}
\mathcal{M} : \mathcal{N} \to \mathcal{T}_\mathrm{a},
\end{equation}
that estimates \( \mathcal{M}^* \).

We assume that for each \( p \in \mathcal{P} \) both the audio timeline and the score timeline are much longer than the number of occurrences \( \mathcal{N}_p \), i.e.,
% \[
% T_\mathrm{a} \gg |\mathcal{N}_p|, \qquad T_\mathrm{s} \gg |\mathcal{N}_p|.
% \]
$T_\mathrm{a},\,T_\mathrm{s} \gg |\mathcal{N}_p|$.

Although \( \mathcal{M}_p^* \) is unknown, we assume access to a coarse onset-level alignment \( \mathcal{M}^\mathrm{seq} : \mathcal{N} \to \mathcal{T}_\mathrm{a} \) induced by a sequence-level alignment (Figure~\ref{fig:overview}, top left), which is locally accurate within a bounded temporal deviation \( B \), independent of \( p \), as in Equation~\ref{eq:bound}.
% , i.e.,
% \begin{equation}
% \bigl| \mathcal{M}^\mathrm{seq}(n) - \mathcal{M}^*(n) \bigr| \le B,
% \quad \forall\, n \in \mathcal{N}.
% \end{equation}

We further assume the existence of a posteriorgram
\begin{equation}
f_\theta^p : \mathcal{T}_\mathrm{a} \to [0,1],
\end{equation}
where \( f_\theta^p(t) \) denotes the likelihood that an onset of pitch \( p \) occurs in the audio at time \( t \). In practice, \( f_\theta^p \) can be obtained from a neural-network-based automatic transcriber pre-trained on a related yet distinct domain.

Our goal is to estimate \( \mathcal{M}_p^* \) using the prior information provided by \( f_\theta^p \) and \( \mathcal{M}^\mathrm{seq} \); that is, we aim to refine the coarse estimation \( \mathcal{M}^\mathrm{seq}_p=\mathcal{M}^\mathrm{seq}|_{\mathcal{N}_p} \) within the admissible windows defined by the temporal error bound \( B \), guided by the posteriorgram \( f_\theta^p \).
For each note occurrence \( n \in \mathcal{N}_p \), 
we define the set of admissible candidate matches as
\begin{equation}
\Omega(n) = 
\bigl\{\, t \in \mathcal{T}_\mathrm{a} \,\bigm|\, 
|t - \mathcal{M}^\mathrm{seq}(n)| \le B \,\bigr\}.
\end{equation}
We refer to \(\Omega(n)\) as the \emph{snapping window}, and to $B$ as the \emph{snapping window size}. Since windows \(\Omega(n)\) may overlap for different $n$, greedy per-window candidate selection (e.g., from left to right) may yield a suboptimal global matching (Figure~\ref{fig:overview}, middle). 
An optimal matching can instead be obtained using
classical bipartite graph matching algorithms~\cite{karp1980algorithm, jonker1987shortest}\footnote{\url{https://docs.scipy.org/doc/scipy/reference/generated/scipy.sparse.csgraph.min_weight_full_bipartite_matching.html}} (Figure~\ref{fig:overview}, right). We define a weighted bipartite graph
% \(
% G = (V_1, V_2, E),
% \)
\begin{equation}
G = (V_1, V_2, E),
\end{equation}
with vertex sets $V_1 = \mathcal{N}_p$ and $V_2 = \mathcal{T}_\mathrm{a}$, and edge set
\begin{equation}
E = \bigcup_{n\in\mathcal{N}_p} \{\, (n, t) \mid t \in \Omega(n) \,\}.
\end{equation}
Each edge \((n, t)\) carries a weight $w_{n, t} = f^p_\theta(t)$, which quantifies the likelihood of matching the onset of the note event \( n \) to audio frame \( t \).
For all \( t \notin \Omega(n) \), we set \( w_{n, t} = 0 \).

We seek the matching
\begin{equation}
\widehat{\mathcal{M}_p} 
= \arg\max_{\mathcal{M}_p}
\sum_{n \in \mathcal{N}_p} 
w_{n, \mathcal{M}_p(n)}\end{equation}
subject to the following constraints:
\begin{equation}
\mathcal{M}_p(n) \in \Omega(n), 
\quad \text{for all } n \in \mathcal{N}_p,
\end{equation}
\begin{equation}
n \ne n' 
\implies 
\mathcal{M}_p(n) \ne \mathcal{M}_p(n'),
\quad \text{for all } n, n' \in \mathcal{N}_p.
\end{equation}
We solve for \(\widehat{\mathcal{M}_p}\) using the algorithms of~\cite{karp1980algorithm, jonker1987shortest}.

% Since snapping windows \(\Omega(n)\) may overlap for different $n$, a greedy local selection may yield a suboptimal global matching (Figure~\ref{fig:overview}, middle). 
% An optimal matching can instead be obtained using classical algorithms for bipartite matching, such as those of~\cite{karp1980algorithm, jonker1987shortest} (Figure~\ref{fig:overview}, right).

As onset-level alignments obtained via snapping are used as labels for training a transcription model, the snapping window size \(B\) directly controls the strength of supervision. It can be adjusted based on the expected accuracy of the initial sequence-level alignment: smaller windows provide stronger supervision but require more precise sequence-level alignment, while larger windows tolerate greater alignment errors at the cost of weaker supervision.

% Since we use the onset-level alignment estimated through snapping as labels for training a transcription model, The snapping window size \(B\) controls the supervision strength. It can be tuned according to our assumptions about the accuracy of the initial sequence-level alignment. A smaller window provides stronger supervision but requires more precise sequence-level alignment, whereas a larger window accommodates greater alignment errors, at the cost of weaker supervision.

% We denote the resulting snapping operator as
% \begin{equation}
% \mathrm{Snap}(\mathcal{N}_p, \mathcal{M}^\mathrm{seq}, B, f^p_\theta) = \widehat{\mathcal{M}}.
% \end{equation}

\section{Instrument-Agnostic Transcription}\label{sec:agnostic}
Instrument-agnostic transcription estimates note activity without distinguishing between instruments. In this section we formalize this task, and outline the simplifying assumptions used in this work.

Starting from a note-event list (Equations~\ref{eq:note_list},~\ref{eq:note_format}), we derive a note-onset piano roll
\(M^\mathrm{on} \in \{0,1\}^{T_\mathrm{s} \times P}\), where \(T_\mathrm{s}\) is the number of timesteps and \(P\) the number of pitch bins:
\begin{equation}
M^\mathrm{on}_{t,p} =
\begin{cases}
1, & \exists\, t^\mathrm{off}\in\mathcal{T}_\mathrm{s},\, i\in\mathcal{I}: (p, t, t^\mathrm{off}, i)\in\mathcal{N},\\
0, & \text{otherwise}.
\end{cases}
\end{equation}
Thus, \(M^\mathrm{on}_{t,p}=1\) if and only if there exists a note of pitch \(p\) that starts at time \(t\).
In this work we focus on onset detection, though a note-activity piano roll considering note duration could be defined similarly.

Because instrument labels are discarded, notes from different instruments with the same pitch and onset time are merged into a single active entry in \(M^\mathrm{on}\). This is an inherent property of the instrument-agnostic formulation.

Our objective is to train a transcriber to directly predict from audio the piano roll corresponding to the underlying (latent) note-event list of the performance.
% Our objective is to train a transcriber that directly predicts this representation from audio, i.e., it predicts the piano roll that would be derived from the underlying (latent) note-event list of the performance.

Training data are obtained by aligning score-derived piano rolls with real performances. In multi-instrument recordings, inconsistencies may arise when notes of the same pitch simultaneous in the score occur at different audio frames, or when notes from different score timesteps coincide in audio. Focusing on chamber music and piano, we assume such cases are rare and do not substantially affect training. They may be more prominent in orchestral settings; however, investigating this phenomenon is left for future work.

\section{Experiments}\label{sec:experiments}
In this section, we present our experiments. We begin by introducing the datasets used for training and evaluation (Section~\ref{sec:datasets}). We then describe our approach for evaluating transcription and alignment accuracy (Section~\ref{sec:eval_types}), followed by cross-dataset transcription evaluation (Section~\ref{sec:musicnet}) and alignment accuracy evaluation (Section~\ref{sec:maestro}).

\subsection{Datasets}\label{sec:datasets}
Below, we describe the datasets used in this work. Importantly, MusicNet~\cite{ThickstunHK17_MusicNet_ICLR} is the \emph{only} dataset used for training; all other datasets---including MAESTRO~\cite{HawthorneSRSHDE19_MAESTRO_ICLR} which provides its own training split---are used solely for evaluation.

\textbf{MusicNet}~\cite{ThickstunHK17_MusicNet_ICLR} is a 34-hour dataset of 11 instruments, including multi-instrument chamber music ensembles of 1--8 players. Its main advantage is acoustic diversity and exclusive use of professional recordings. Its main limitation is low alignment accuracy of score annotations, making them insufficient for training transcription systems. We correct annotation timing errors using snapping, which we show to be highly effective.
% \textbf{MusicNet}~\cite{ThickstunHK17_MusicNet_ICLR} is a 33-hour dataset featuring 12 different instruments recorded in diverse acoustic settings, including multi-instrument chamber music performances involving ensembles of 1--8 players. Its primary advantage is its acoustic diversity and the exclusive use of professional recordings. Its main limitation is the low alignment accuracy of score annotations, making them insufficient for training transcription systems. We address the annotation timing errors using snapping, which we show to be highly effective.

\textbf{MAESTRO}~\cite{HawthorneSRSHDE19_MAESTRO_ICLR} is a 140-hour dataset of solo piano performances recorded on a Disklavier. It provides highly accurate onset and offset annotations,
% on the order of 3\,ms,
but is limited in acoustic diversity. We use only the 20-hour test set for evaluation, disregarding the training set.

\textbf{Saarland Music Data (SMD)}~\cite{MuellerKBA11_SMD_ISMIR-lateBreaking} is a 5-hour dataset of solo piano performances recorded on a Disklavier, similar to MAESTRO but offering different acoustic conditions.

\textbf{URMP}~\cite{LiLDDS19_MultitrackDataset_TMM} is a multi-instrument, multi-track chamber music dataset totaling 80 minutes. Each musical piece comprises up to five instrumental tracks. Its isolated monophonic recordings enable relatively accurate onset annotations (although not as precise as those of a Disklavier). Nevertheless, the dataset is small and acoustically uniform.
% , as all tracks were recorded in the same environment.

\textbf{ChoraleBricks}~\cite{BalkeBM25_ChoraleBricks_TISMIR} is a multi-track wind music dataset consisting of four-part chorales, created using methods similar to those employed for URMP. The total duration of all pieces is less than 7 minutes. However, each of the four parts in every piece is recorded with multiple instruments, which gives the possibility of creating many combinations of instruments for each chorale. For our work, we use all possible four-part combinations, resulting in over 52 hours.
% of music.

% \textbf{PHENICX}~\cite{LiemGS15_PHENICX_ICMWE, SchedlHTML16_PHENICX-SMM_CBMI}
\textbf{PHENICX}~\cite{LiemGS15_PHENICX_ICMWE}
is an orchestral multi-track dataset created similarly to URMP and ChoraleBricks, but featuring full orchestral performances and a total duration of just over 10 minutes. Each piece comprises between 10 and 40 instrumental tracks. Like URMP, onset annotations are relatively accurate, but the dataset lacks acoustic diversity due to its uniform recording environment.

\textbf{The Beethoven Symphony Excerpts Dataset (BSED)} is an internal orchestral evaluation dataset with a total duration of 37 minutes, created by aligning musical scores with corresponding audio recordings. Its primary advantages are its acoustic diversity and the professional quality of the recorded performances. As separated tracks are not available, annotations were produced using audio--score alignment techniques~\cite{EwertMG09_HighResAudioSync_ICASSP, MuellerOKPD21_SyncToolbox_JOSS}. 
% Consequently, temporal annotation errors may be higher than for the aforementioned multi-track datasets, which we account for appropriately.
% ;
% we account for this appropriately in our evaluation.

% \textbf{The Beethoven Symphony Excerpts Dataset (BSED)} is an \ben{internal} orchestral evaluation dataset with a total duration of 37 minutes, created by aligning musical scores with corresponding audio recordings. Its primary advantage is its acoustic diversity. As separated tracks are not available, annotations were produced using audio-score alignment techniques~\cite{MuellerOKPD21_SyncToolbox_JOSS, OezerKM21_SyncToolbox_ISMIR-LBD}. Since score--audio alignment was performed on existing mixed recordings, temporal errors in annotation may be higher than for the aformentioned multi-track datasets. We account for this appropriately in our evaluation.
% that combine signal processing and neural features.

\subsection{Evaluating Alignment Accuracy}\label{sec:eval_types}
There are two general approaches to evaluating alignment accuracy. The first is direct, comparing the estimated alignment to a ground-truth reference. The second is indirect, measuring how alignment affects transcription performance when aligned scores serve as training labels. The direct approach is explicit but less practical, as reference alignments at millisecond precision are rare. The indirect approach, though less explicit, is more scalable and provides meaningful evaluation via its effect on transcription metrics.
% There are two general approaches to evaluating alignment accuracy. The first is direct, comparing the estimated alignment to a ground-truth reference. The second is indirect, measuring how alignment affects transcription performance when aligned scores are used as training labels. While the direct approach is explicit, it is less practical since reference alignments with millisecond-precision are rare. The indirect approach, though less explicit, is more scalable and provides meaningful evaluation through the impact on transcription metrics.

Accordingly, we adopt the indirect approach as our primary evaluation, measuring alignment quality through its effect on transcription performance. To complement this, we also perform a direct evaluation in a controlled setting. Our experiments correspond to the two approaches:

(i) \textbf{Cross-dataset transcription (Section~\ref{sec:musicnet}):} We compare how different alignment policies used in training data annotation affect transcription performance.

(ii) \textbf{Direct alignment evaluation  (Section~\ref{sec:maestro}):} Using the MAESTRO dataset which has high-accuracy reference alignments, we test alignment quality by attempting to recover the ground truth from intentionally perturbed labels.

\begin{figure}
    \centering
    \includegraphics[width=\linewidth]{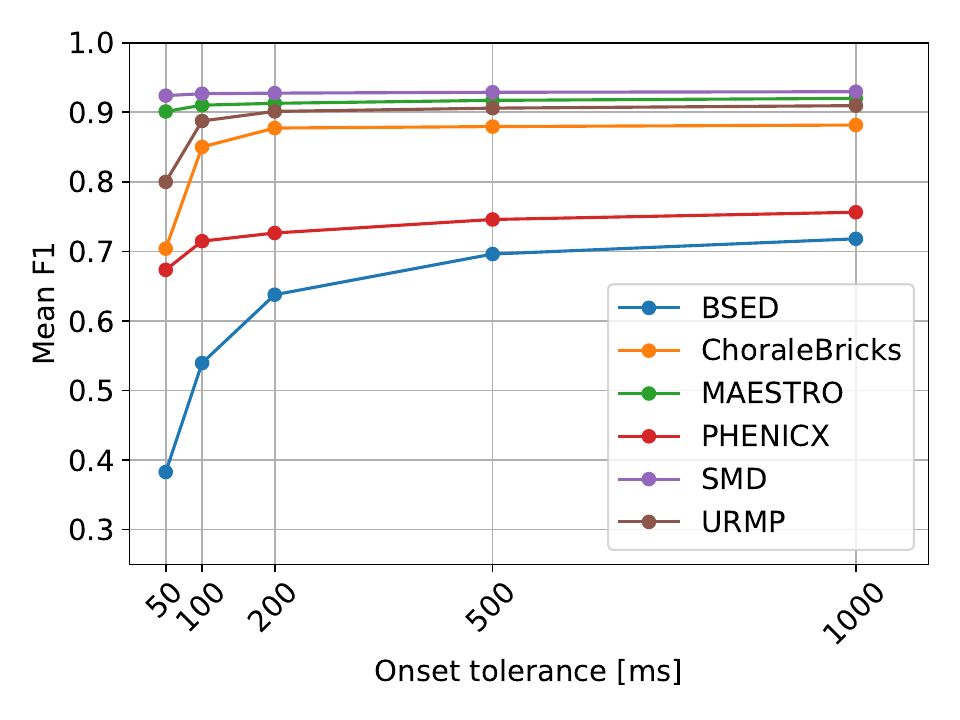}
    \caption{Note-level F1 score as a function of onset tolerance threshold, using the \texttt{DTW-BiP} (0.64s) model.}
    \label{fig:f1_over_tol}
\end{figure}

\subsubsection{Metrics}\label{sec:metrics}
Our evaluation metrics consist of note-level precision, recall, and F1 score, where a predicted note is counted as correct if its onset is within a temporal threshold of the reference onset timing. In the following section (\ref{sec:misalignment}) we elaborate on the exact threshold used.
% We use the standard 50\,ms threshold for datasets with high-accuracy reference alignments (MAESTRO, SMD). For datasets that may contain annotation errors in onset timing (URMP, ChoraleBricks, PHENICX, BSED), we apply a more lenient threshold---100\,ms in all cases except BSED---as detailed in the following section (\ref{sec:misalignment}).
% In addition, we report the strict 50\,ms–threshold results for all datasets.

\subsubsection{Handling Evaluation Set Misalignment}\label{sec:misalignment}
It is important to recognize that some datasets may contain timing inaccuracies in their annotations, particularly those not recorded with specialized hardware such as a Disklavier. To prevent distorted evaluation, we adjust the onset tolerance according to the expected annotation quality.

For the piano datasets MAESTRO and SMD, the Disklavier provides temporal precision on the order of 3 ms. Consequently, for these datasets we adopt the standard 50 ms onset-tolerance threshold.

In the strings and winds datasets---URMP, ChoraleBricks, PHENICX, and BSED\footnote{In addition to sequence-level alignments, BSED also provides accurate note-level alignments obtained using snapping. Since this evaluation investigates snapping itself, we intentionally do not use them. Furthermore, for similar reasons, we use sequence-level alignments derived from signal-processing features rather than neural transcription features.}---note-onset and offset annotations may be misaligned. In principle, these misalignments could be corrected using snapping; however, because snapping is precisely the method we aim to evaluate, doing so would bias the results. To accommodate small alignment errors, we instead apply a 100~ms onset tolerance threshold for URMP, ChoraleBricks, and PHENICX.
% In strings and winds datasets---URMP, ChoraleBricks, PHENICX, and BSED---note-onset and offset annotations may be misaligned.\footnote{BSED also includes accurate note-level alignments obtained using snapping, which we intentionally do not use for this evaluation, which evaluates snapping itself.} In principle,
% this could be corrected using snapping;
% however, because snapping is precisely the method we intend to evaluate, doing so would bias the results. Instead, to accommodate small alignment errors, we use a 100 ms onset tolerance threshold for URMP, ChoraleBricks, and PHENICX.

As shown in Figure \ref{fig:f1_over_tol}, this results in a substantial increase in F1 for URMP, ChoraleBricks, PHENICX, whereas the impact on MAESTRO and SMD—whose annotations are highly accurate—is minimal. This pattern suggests that lower performance under tight thresholds often reflects annotation errors rather than transcription errors. We therefore consider 100 ms a reasonable approximation of typical human annotation error for these datasets.

% The labels we use are obtained via score--audio DTW-based sequence-level alignment performed on full polyphonic mixes, without any snapping step, which can introduce substantially larger timing errors.
% , and another produced by applying an additional snapping step.
% Because our goal is to assess snapping itself,
% we use the DTW-aligned labels \emph{without} snapping to ensure an unbiased evaluation.
% \textbf{BSED} labels are obtained via DTW-based sequence alignment\footnotemark[\value{footnote}] performed on full polyphonic mixes, which can introduce larger timing errors. To account for sequence-level alignment inaccuracies, which may reach 0.5--1 seconds as demonstrated in the following sections, we use a tolerance threshold of 500\,ms. Figure~\ref{fig:f1_over_tol} supports this choice: whereas in the multitrack datasets most of the F1 increase occurs when raising the tolerance from 50\,ms to 100\,ms, the F1 scores for BSED increase substantially up to a 500\,ms tolerance.
\textbf{BSED} labels are obtained via DTW-based sequence alignment\footnotemark[\value{footnote}] on full polyphonic mixes, which can introduce larger timing errors. To account for sequence-level alignment inaccuracies, which may reach 0.5--1 seconds as shown in the following sections, we use a tolerance threshold of 500\,ms. Figure~\ref{fig:f1_over_tol} supports this choice: whereas in the multitrack datasets most of the F1 increase occurs when raising the tolerance from 50\,ms to 100\,ms, the F1 scores for BSED increase substantially up to a 500\,ms tolerance.

\subsection{Cross-dataset Transcription: MusicNet}\label{sec:musicnet}
In this section we focus exclusively on cross-dataset evaluation to better reflect real-world conditions: All compared models are trained on MusicNet, and tested on MAESTRO, SMD, URMP, ChoraleBricks, PHENICX, and BSED.
All models are initialized from the same instrument-agnostic synthetic pre-training, released by~\cite{YaffeMMB25_CountEM_ISMIR}.
% In this section---which is the core of our experiments---we compare alignment strategies by examining how they affect transcription performance when used for labeling. We focus exclusively on cross-dataset evaluation to better reflect real-world conditions: All compared models are trained on MusicNet, and tested on six external datasets spanning a range of ensemble complexities. These include two piano datasets (MAESTRO, SMD), two small-ensemble datasets featuring strings and winds (URMP, ChoraleBricks), and two full-orchestra datasets (PHENICX, BSED). All models are initialized from the same instrument-agnostic synthetic pre-training, released by~\cite{YaffeMMB25_CountEM_ISMIR}.

We report results in order of increasing ensemble complexity. We begin with piano transcription (Section~\ref{sec:piano}), proceed to instrument-agnostic transcription for small string and wind ensembles (Section~\ref{sec:small_ensemble}), and finally address instrument-agnostic orchestral transcription (Section~\ref{sec:orchestra}).

\subsubsection{Compared Models}
We compare models trained in an EM manner~\cite{MamanBermano22_UnalignedAMT_ICML} under different alignment strategies varying in snapping algorithm, window size, and initial sequence alignment. We consider two snapping policies: Our proposed bipartite graph matching (\texttt{BiP}) and greedy snapping (\texttt{Gre}), commonly used in prior work. Snapping windows range from 0.1\,s to 60\,s.
% We compare models trained in an EM manner~\cite{MamanBermano22_UnalignedAMT_ICML} under different alignment strategies varying in the snapping algorithm, window size, and initial sequence alignment. Specifically, we consider two snapping policies: Our proposed bipartite graph matching (\texttt{BiP}), and greedy snapping (\texttt{Gre}), commonly used in prior work. Snapping windows range from 0.1\,s to 60\,s.

Two types of initial sequence-level alignments are considered: Dynamic Time Warping based on chroma and onset features~\cite{EwertMG09_HighResAudioSync_ICASSP} (\texttt{DTW}) and linear stretching of the score to the audio timeline (\texttt{LS}). For \texttt{DTW}, snapping windows range from 0.1\,s to 2\,s. Since \texttt{LS} can introduce substantial alignment errors, it requires larger snapping windows to compensate; accordingly, we use windows of 2\,s to 60\,s.
% Two types of initial sequence-level alignments are considered: Dynamic Time Warping based on chroma and onset features~\cite{EwertMG09_HighResAudioSync_ICASSP} (\texttt{DTW}) and linear stretching of the score to the audio timeline (\texttt{LS}). For \texttt{DTW}, snapping windows range from 0.1\,s to 2\,s. Since \texttt{LS} can introduce substantial alignment errors, it requires larger snapping windows to compensate; accordingly, we use windows ranging from 2\,s to 60\,s.

For example, \texttt{DTW-BiP} (0.64s) denotes bipartite graph-based snapping with 0.64-second windows applied after DTW, whereas \texttt{LS-Gre} (60s) denotes greedy snapping with 60-second windows applied after linear stretching.

We also compare our method to a recent top-$K$ peak-picking histogram-based approach~\cite{YaffeMMB25_CountEM_ISMIR} (\texttt{Hist}).

Finally, \texttt{Synth} refers to the instrument-agnostic, synthetically pre-trained model released by~\cite{YaffeMMB25_CountEM_ISMIR}, from which all compared models are initialized.

\subsubsection{Piano Transcription}\label{sec:piano}
% \textbf{P} & \textbf{R} & \textbf{F1}

\begin{table}[t!]
\centering
\setlength{\tabcolsep}{4pt}
\begin{tabular}{lccc|ccc}
\toprule
\multirow{2}{*}{\textbf{Transcriber}} & \multicolumn{3}{c}{\textbf{MAESTRO}} & \multicolumn{3}{c}{\textbf{SMD}} \\
\cmidrule(lr){2-4}
\cmidrule(lr){5-7}
& \textbf{P} & \textbf{R} & \textbf{F1} & \textbf{P} & \textbf{R} & \textbf{F1} \\
\cmidrule(lr){2-7}
\texttt{Synth} & 88.4 & 81.6 & 84.7 & 93.0 & 85.6 & 88.9 \\
\texttt{DTW} & 96.6 & 48.6 & 62.1 & 96.1 & 56.0 & 69.0 \\
\midrule
% \rowcolor{gray!7}
\multicolumn{7}{c}{\textbf{Snapping: Previous Work~\cite{MamanBermano22_UnalignedAMT_ICML, YaffeMMB25_CountEM_ISMIR}}} \\
\midrule
\texttt{DTW-Gre} (2s) & 94.0 & 84.8 & 89.0 & 97.5 & 87.4 & 92.1 \\
\texttt{DTW-Gre} (0.64s) & 93.8 & 85.2 & 89.2 & 97.5 & 87.2 & 91.9 \\
\texttt{DTW-Gre} (0.1s) & 95.8 & 84.5 & 89.6 & 98.1 & 85.9 & 91.4 \\
\texttt{LS-Gre} (60s) & 92.8 & 79.0 & 85.2 & 95.1 & 84.5 & 89.3 \\
\texttt{Hist} & 94.6 & 86.0 & 89.9 & 97.6 & 88.1 & 92.5 \\
\midrule
% \rowcolor{gray!7}
\multicolumn{7}{c}{\textbf{Snapping: Ours}} \\
\midrule
\texttt{DTW-BiP} (2s) & 94.4 & \textbf{86.6} & \textbf{90.2} & 97.9 & \textbf{88.2} & \textbf{92.7} \\
\texttt{DTW-BiP} (0.64s) & 95.0 & 85.9 & 90.1 & \textbf{98.2} & 87.5 & 92.4 \\
\texttt{DTW-BiP} (0.1s) & \textbf{96.3} & 84.6 & 89.9 & 98.1 & 86.9 & 92.0 \\
\texttt{LS-BiP} (60s) & 93.1 & 84.2 & 88.3 & 96.9 & 87.6 & 91.9 \\
\texttt{LS-BiP} (20s) & 94.5 & 83.1 & 88.2 & 97.8 & 86.6 & 91.7 \\
\texttt{LS-BiP} (10s) & 94.4 & 80.4 & 86.6 & 97.7 & 85.7 & 91.1 \\
\texttt{LS-BiP} (2s) & 94.5 & 52.4 & 65.1 & 96.1 & 64.0 & 74.4 \\
\bottomrule
\end{tabular}
\caption{Piano transcription results. Models are trained on MusicNet and evaluated on MAESTRO and SMD.}\label{table:piano}
\end{table}

Table~\ref{table:piano} presents piano transcription results, training on MusicNet and evaluating on MAESTRO and SMD.

As shown, \texttt{DTW} alone produces poor labels, substantially reducing F1 compared to the \texttt{Synth} baseline (e.g., MAESTRO: 84.7\% $\rightarrow$ 62.1\%). Applying snapping after \texttt{DTW} yields large gains ($\sim$23--28\%), increasing MAESTRO F1 to 89--90\%. Relative to \texttt{Synth}, snapping provides a $\sim$3--5\% improvement (84.7\% $\rightarrow$ 90.2\% with \texttt{DTW-BiP}, 0.64\,s), demonstrating effective domain adaptation with weakly aligned data. These results show that snapping is essential for training with weakly aligned labels and that it makes such training beneficial.
% As can be seen, \texttt{DTW} alone produces poor labels, reducing F1 compared to the \texttt{Synth} baseline (e.g., MAESTRO: 84.7\(\to\)62.1\%). applying snapping after \texttt{DTW} yields large gains (\(\sim\)23–28\%), boosting MAESTRO F1 to 89–90\%. Compared to \texttt{Synth}, snapping improves performance by \(\sim\)3–5\% (84.7\(\to\)90.2\% with \texttt{DTW-BiP} 0.64s), demonstrating effective domain adaptation with weakly-aligned data. This shows that snapping is essential for training with weakly-aligned labels.

After showing snapping to be essential, we compare different snapping methods. Our graph-based approach yields marginal yet consistent improvements of approximately 0.5--1.2\% across models and datasets over greedy snapping used in prior work~\cite{MamanBermano22_UnalignedAMT_ICML}. For example, on MAESTRO, using a 2\,s window, F1 increases from 89.0 (\texttt{DTW-Gre} 2s) to 90.2 (\texttt{DTW-BiP} 2s). We further observe that the recent top-$K$ peak picking histogram-based method (\texttt{Hist}) slightly outperforms greedy snapping. For example, on SMD, the F1 score increases from 91.9 (\texttt{DTW-Gre} 0.64s) to 92.5 (\texttt{Hist}). We hypothesize that this is because the top-$K$ method does not suffer from issues caused by overlapping snapping windows. However, graph-based snapping (e.g., \texttt{DTW-BiP} 2s) still slightly outperforms the histogram top-$K$ method. Although the difference is small for piano transcription, we will see that the gap becomes larger for strings and winds, reaching up to 5\% (Sections~\ref{sec:small_ensemble},~\ref{sec:orchestra}).

Next, we analyze the effect of snapping window size. A 0.1\,s window already substantially outperforms the \texttt{Synth} baseline and \texttt{DTW}, and larger windows yield further gains, mainly through higher recall. On SMD, increasing the window from 0.1\,s to 2\,s raises recall and F1 from $86.9/92.0$ to $88.2/92.7$ (\texttt{DTW-BiP} 2s).

Finally, we examine extremely large windows. With linear stretching, small windows fail (F1 on MAESTRO degrades from $84.7$ to 65.1 with \texttt{LS-BiP} $2\,\mathrm{s}$). Compensating with 60-second windows,
% However, \texttt{LS} with very large windows ($60\,\mathrm{s}$) shows a clear gap:
greedy snapping (\texttt{LS-Gre} $60\,\mathrm{s}$) yields minimal gains if any at all compared to \texttt{Synth} (F1 $84.7 \to 85.2$, recall $81.6 \to 79.0$), whereas graph-based (\texttt{LS-BiP} $60\,\mathrm{s}$) substantially improves both recall and F1 (F1 $84.7 \to 88.3$, recall $81.6 \to 84.2$). This shows that as the window size grows, greedy snapping degrades due to increased overlaps, while graph-based snapping remains robust.

\subsubsection{Strings \& Winds Transcription---Small Ensemble}\label{sec:small_ensemble}
\begin{table}[t!]
\centering
\setlength{\tabcolsep}{4pt}
\begin{tabular}{lccc|ccc}
\toprule
\multirow{2}{*}{\textbf{Transcriber}} & \multicolumn{3}{c}{\textbf{URMP}} & \multicolumn{3}{c}{\textbf{ChoraleBricks}} \\
\cmidrule(lr){2-4}
\cmidrule(lr){5-7}
 & \textbf{P} & \textbf{R} & \textbf{F1} & \textbf{P} & \textbf{R} & \textbf{F1} \\
\cmidrule(lr){2-7}
\texttt{Synth} & 77.6 & 78.0 & 77.5 & 87.5 & 72.7 & 79.1 \\
\texttt{DTW} & 97.1 & 62.9 & 75.5 & 97.3 & 53.9 & 68.6 \\
\midrule
% \rowcolor{gray!10}
\multicolumn{7}{c}{\textbf{Snapping: Previous Work~\cite{MamanBermano22_UnalignedAMT_ICML, YaffeMMB25_CountEM_ISMIR}}} \\
\midrule
\texttt{DTW-Gre} (2s) & 91.2 & 84.5 & 87.5 & 93.3 & 77.0 & 83.9 \\
\texttt{DTW-Gre} (0.64s) & 93.0 & 85.5 & 89.0 & 94.2 & 80.1 & 86.2 \\
\texttt{DTW-Gre} (0.1s) & \textbf{94.2} & 83.7 & 88.4 & \textbf{95.2} & 77.3 & 84.9 \\
\texttt{LS-Gre} (60s) & 83.3 & 78.4 & 80.3 & 81.1 & 78.4 & 78.9 \\
\texttt{Hist} & 88.7 & 85.7 & 87.1 & 90.1 & 79.0 & 83.7 \\
\midrule
% \rowcolor{gray!10}
\multicolumn{7}{c}{\textbf{Snapping: Ours}} \\
\midrule
\texttt{DTW-BiP} (2s) & 90.8 & \textbf{86.8} & 88.6 & 91.7 & \textbf{82.4} & 86.5 \\
\texttt{DTW-BiP} (0.64s) & 92.8 & 86.3 & \textbf{89.3} & 93.4 & 82.2 & \textbf{87.0} \\
\texttt{DTW-BiP} (0.1s) & 93.4 & 85.2 & 88.9 & 94.9 & 77.5 & 84.8 \\
\texttt{LS-BiP} (60s) & 87.3 & 81.4 & 84.0 & 89.1 & 73.0 & 79.2 \\
\texttt{LS-BiP} (20s) & 85.8 & 83.5 & 84.4 & 89.1 & 75.6 & 80.9 \\
\texttt{LS-BiP} (10s) & 90.1 & 81.1 & 85.2 & 92.6 & 70.1 & 78.7 \\
\texttt{LS-BiP} (2s) & 88.8 & 72.9 & 79.2 & 91.9 & 63.6 & 74.6 \\
\bottomrule
\end{tabular}
\caption{Transcription results on small multi-instrument ensembles. Models are trained on MusicNet and evaluated on URMP and ChoraleBricks.}\label{table:small_ensemble}
\end{table}

Table~\ref{table:small_ensemble} presents results for \emph{instrument-agnostic} (multi-instrument) transcription on small musical ensembles. The compared models were trained on MusicNet and evaluated on URMP (chamber music with strings and winds) and ChoraleBricks (small wind ensembles).

Similar trends are observed to those seen in piano transcription: \texttt{DTW} labels degrade performance relative to the \texttt{Synth} baseline. For example, F1 on ChoraleBricks drops from 79.1 to 68.6.
Applying snapping after \texttt{DTW} substantially improves F1. For example, \texttt{DTW-BiP} with a 0.64 s window increases F1 relative to \texttt{Synth} from 77.5 to 89.3 on URMP and from 79.1 to 87.0 on ChoraleBricks.

Graph-based snapping outperforms both greedy snapping and the histogram-based method. With 2\,s windows, graph-based snapping (\texttt{DTW-BiP} 2s) improves F1 over greedy snapping (\texttt{DTW-Gre} 2s) from 83.9 to 86.5 on ChoraleBricks and from 87.5 to 88.6 on URMP. Compared to \texttt{Hist}, F1 increases from 83.7 to 86.5 on ChoraleBricks and from 87.1 to 88.6 on URMP. Larger windows amplify the benefit of graph-based snapping; on URMP, \texttt{LS-BiP} 60s improves F1 relative to \texttt{LS-Gre} 60s from 80.3 to 84.4.

Overall, results on URMP slightly exceed those on ChoraleBricks, possibly due to the higher proportion of string instruments in MusicNet and the higher annotation precision of URMP; increasing the onset tolerance from 100\,ms to 200\,ms raises F1 by $\sim$3\% on ChoraleBricks, but only by $\sim$1\% on URMP (Figure~\ref{fig:f1_over_tol}).

Compared to piano results (Table~\ref{table:piano}), performance on URMP and ChoraleBricks falls within a similar range, indicating that onset detection can be effective for string and wind instruments despite their less pronounced onsets.

\begin{table}[t!]
\centering
\setlength{\tabcolsep}{4pt}
\begin{tabular}{lccc|ccc}
\toprule
\multirow{2}{*}{\textbf{Transcriber}} & \multicolumn{3}{c}{\textbf{PHENICX}} & \multicolumn{3}{c}{\textbf{BSED}} \\
\cmidrule(lr){2-4}
\cmidrule(lr){5-7}
 & \textbf{P} & \textbf{R} & \textbf{F1} & \textbf{P} & \textbf{R} & \textbf{F1} \\
\cmidrule{2-7}
\texttt{Synth} & 83.8 & 54.8 & 66.0 & 83.2 & 39.5 & 51.3 \\
\texttt{DTW} & 93.7 & 41.9 & 57.2 & 89.3 & 15.4 & 24.3 \\
\midrule
\multicolumn{7}{c}{\textbf{Snapping: Previous Work~\cite{MamanBermano22_UnalignedAMT_ICML, YaffeMMB25_CountEM_ISMIR}}} \\
\midrule
\texttt{DTW-Gre} (2s) & 85.8 & 60.0 & 70.4 & 81.6 & 59.9 & 67.2\\
\texttt{DTW-Gre} (0.64s) & 85.5 & 60.3 & 70.4 & 78.9 & 62.4 & 67.6 \\
\texttt{DTW-Gre} (0.1s) & 87.1 & 60.6 & 71.2 & 83.9 & 53.4 & 62.3\\
\texttt{LS-Gre} (60s) & 78.7 & 62.2 & 68.8 & 72.3 & 59.1 & 62.4\\
\texttt{Hist} & 86.3 & 60.7 & 71.2 & 74.8 & 58.4 & 64.1 \\

\midrule
\multicolumn{7}{c}{\textbf{Snapping: Ours}} \\
\midrule
\texttt{DTW-BiP} (2s) & 86.3 & 61.5 & 71.5 & 79.7 & 62.2 & 67.8\\
\texttt{DTW-BiP} (0.64s) & 84.4 & \textbf{62.3} & 71.5 & 81.4 & \textbf{63.8} & \textbf{69.6} \\
\texttt{DTW-BiP} (0.1s) & \textbf{87.6} & 60.9 & \textbf{71.7} & \textbf{84.4} & 53.9 & 63.4 \\
\texttt{LS-BiP} (60s) & 85.1 & 61.4 & 71.0 & 72.6 & 62.2 & 64.7 \\
\texttt{LS-BiP} (20s) & 84.9 & 60.7 & 70.5 & 73.3 & 62.0 & 64.9 \\
\texttt{LS-BiP} (10s) & 85.4 & 60.3 & 70.3 & 73.3 & 61.3 & 64.5 \\
\texttt{LS-BiP} (2s) & 85.6 & 36.4 & 50.8 & 74.1 & 47.4 & 55.0 \\
\bottomrule
\end{tabular}
\caption{Orchestral transcription results. Models are trained on MusicNet and evaluated on BSED and PHENICX.}\label{table:orchestra}
\end{table}

% \begin{table}[t!]
% \centering
% \setlength{\tabcolsep}{4pt}
% \begin{tabular}{lccc|ccc}
% \toprule
% \multirow{2}{*}{Transcriber} & \multicolumn{3}{c}{BSED-DTW} & \multicolumn{3}{c}{BSED} \\
% \cmidrule{2-7}
%  & P & R & F1 & P & R & F1 \\
% \midrule
% \texttt{Synth} & 83.2 & 39.5 & 51.3 & 77.0 & 36.7 & 47.8 \\
% \texttt{DTW} & 89.3 & 15.4 & 24.3 & 86.7 & 15.0 & 23.7 \\
% \midrule
% & \multicolumn{3}{c}{Previous Work} \\
% \midrule
% \texttt{DTW-Gr} (2s) & 81.6 & 59.9 & 67.2 & 76.7 & 56.6 & 63.4 \\
% \texttt{DTW-Gr} (0.640s) & 78.9 & 62.4 & 67.6 & 73.4 & 58.6 & 63.4 \\
% \texttt{DTW-Gr} (0.096s) & 83.9 & 53.4 & 62.3 & 79.6 & 50.7 & 59.3 \\
% \texttt{LS-Gr} (60s) & 72.3 & 59.1 & 62.4 & 68.6 & 56.0 & 59.2 \\
% \texttt{Hist} & 74.8 & 58.4 & 64.1 & 69.1 & 54.4 & 59.5 \\
% \midrule
% & \multicolumn{3}{c}{Ours} \\
% \midrule
% \texttt{DTW-Bp} (2s) & 79.7 & 62.2 & 67.8 & 74.6 & 58.5 & 63.7 \\
% \texttt{DTW-Bp} (0.64s) & 81.4 & 63.8 & 69.6 & 75.6 & 59.6 & 64.9 \\
% \texttt{DTW-Bp} (0.096s) & 84.4 & 53.9 & 63.4 & 79.0 & 51.0 & 59.8 \\
% \texttt{LS-Bp} (60s) & 72.6 & 62.2 & 64.7 & 68.8 & 59.1 & 61.5 \\
% \texttt{LS-Bp} (20s) & 73.3 & 62.0 & 64.9 & 69.0 & 58.8 & 61.4 \\
% \texttt{LS-Bp} (10s) & 73.3 & 61.3 & 64.5 & 69.6 & 58.6 & 61.6 \\
% \texttt{LS-Bp} (2s) & 74.1 & 47.4 & 55.0 & 68.9 & 44.5 & 51.6 \\

% \bottomrule
% \end{tabular}
% \caption{Results for BSED-DTW with 500ms onset tolerance and BSED-Note with 100ms tolerance.}\label{table:orchestra}
% \end{table}

\subsubsection{Orchestral Transcription}\label{sec:orchestra}

Table~\ref{table:orchestra} shows orchestral transcription results, traning on MusicNet and evaluating on PHENICX and BSED. As with piano and small ensembles, \texttt{DTW} alone is insufficient, while snapping (\texttt{DTW-BiP}, \texttt{DTW-Gre}) improves alignment, with graph-based snapping consistently outperforming greedy snapping. For example, \texttt{DTW-BiP} 0.64s raises F1 on PHENICX compared to \texttt{DTW-Gre} 0.64s from 70.4 to 71.5 and on BSED from 67.6 to 69.6. Improvement on BSED compared to \texttt{Hist} is substantial: From 64.1 to 69.6. 

Overall metrics are lower than for piano or small-ensemble settings. For example, \texttt{DTW-BiP} (0.64\,s) achieves an F1 score of 89.3 on URMP but only 71.7 on PHENICX. We attribute this gap to increased orchestral complexity: Performances in PHENICX involve 10--40 instruments, while URMP ensembles include no more than five.

\subsection{Alignment Evaluation---MAESTRO}\label{sec:maestro}
We evaluate alignment accuracy on the MAESTRO dataset, for which precise reference annotations are available. We randomly and uniformly perturb the MAESTRO onset labels within a window $[-w, w]$, where $w \in \{1, 5, 15, 60\}$ frames, corresponding to $0.032\,\mathrm{s}$, $0.16\,\mathrm{s}$, $0.48\,\mathrm{s}$, and $1.92\,\mathrm{s}$. The perturbed onsets are then recovered using snapping with the same window size. Alignment quality is measured by the F1 score with respect to the ground truth.
% In this section, we evaluate alignment accuracy on the MAESTRO dataset, for which precise reference annotations are available. We randomly and uniformly perturb the MAESTRO onsets within a window $[-w, w]$, where $w\in\{1, 5, 15,  60\}$ frames, corresponding to $0.032\,\mathrm{s}$, $0.16\,\mathrm{s}$, $0.48\,\mathrm{s}$, and $1.92\,\mathrm{s}$. The perturbed onsets are then recovered using snapping with the same window size. Alignment quality is measured by the F1 score with respect to the ground truth.

We compare three posteriorgrams obtained from the following: (i) the synthetically pre-trained model (\texttt{Synth}); (ii) the \texttt{DTW-BiP} 0.64\,s model fine-tuned from \texttt{Synth} on MusicNet with weakly-aligned labels using snapping (\texttt{MN}); and (iii) the ground-truth piano roll (\texttt{GT}) as a sanity check.

Results are shown in Table~\ref{table:perturbation}. The \texttt{GT} posteriorgram achieves an F1 score of 100\% for all \(w\), for both greedy and graph-based snapping, confirming that snapping operates correctly with ideal posteriorgrams.

With \texttt{Synth}, F1 decreases as \(w\) increases, but graph-based snapping (\texttt{BiP}) still remains robust, achieving F1 of 91.0 at \(w=0.48\,\mathrm{s}\) and 87.8 at \(w=1.92\,\mathrm{s}\), compared to F1 scores of 16.7 and 9.6 for the perturbed labels (\texttt{Pert}). The differences between graph-based and greedy snapping grow with \(w\): 93.6--92.5 = 1.1 for \(w=0.16\,\mathrm{s}\), 91.0--88.6 = 2.4 for \(w=0.48\,\mathrm{s}\), and 87.8--84.7 = 3.1 for \(w=1.92\,\mathrm{s}\). This can be expected: larger windows increase overlap, making greedy snapping less optimal.

The MusicNet-trained model (\texttt{MN}) shows similar trends. However, it has consistently improved performance across snapping methods and window sizes compared to \texttt{Synth}; for example, at \(w=0.48\,\mathrm{s}\) with graph-based snapping (\texttt{BiP}), F1 increases from 91.0 (\texttt{Synth}) to 92.8 (\texttt{MN}). This demonstrates that training with snapping on weakly-aligned data not only improves transcription metrics (as seen in Section~\ref{sec:musicnet}), but also enhances alignment accuracy when using transcription features for snapping.

Finally, comparing with piano transcription results in Table~\ref{table:piano}, we find that the difference between greedy and graph-based snapping is larger for alignment than for transcription (up to 4\% versus about 1\%). This may indicate that the training process is robust to some degree of label noise, provided that a large portion of onsets is accurately aligned.

\begin{table}[t!]
\centering
\begin{tabular}{lc|c|ccc}
\toprule
%\textbf{Method} & \(w\) [s] & \texttt{Pert} & \texttt{Synth} & \texttt{MN} & \texttt{GT} \\
\textbf{Method} & \textbf{\(\boldsymbol{w}\) [s]} & \texttt{Pert} & \texttt{Synth} & \texttt{MN} & \texttt{GT} \\
\midrule
\multirow{4}{*}{\texttt{Gre}}
 & 0.032 & 100  & 95.9 & 96.4 & 100 \\
 & 0.16  & 30.1 & 92.5 & 93.0 & 100 \\
 & 0.48  & 16.7 & 88.6 & 89.9 & 100 \\
 & 1.92  & 9.6  & 84.7 & 87.1 & 100 \\
\midrule
\multirow{4}{*}{\texttt{BiP}}
 & 0.032 & 100  & 95.9 & 96.4 & 100 \\
 & 0.16  & 30.1 & 93.6 & 94.3 & 100 \\
 & 0.48  & 16.7 & 91.0 & 92.8 & 100 \\
 & 1.92  & 9.6  & 87.8 & 91.1 & 100 \\
\bottomrule
\end{tabular}
\caption{F1 score evaluation of snapping applied to the perturbed MAESTRO test set (\texttt{Pert}) across varying perturbation windows \(w\), using different onset posteriorgrams (\texttt{Synth}, \texttt{MN}, \texttt{GT}).}
\label{table:perturbation}
\vspace{-0.07cm}
\end{table}

\section{Conclusion}\label{sec:discussion}
In this work, we investigated \emph{snapping}---the refinement of sequence-level alignments into precise onset-level alignments using neural onset posteriorgrams. Through cross-dataset evaluations and controlled experiments, we showed that snapping is highly effective for onset-level alignment, enabling training transcribers with weakly-aligned labels with further gains when following a graph-based approach. Although we focused on instrument-agnostic AMT, the methods used may be extended to instrument-sensitive transcription, and possibly other timing-critical MIR tasks such as drum transcription or multi-pitch estimation.

\begin{acknowledgments}
This work was funded by the Deutsche Forschungsgemeinschaft (DFG, German Research Foundation) under Grant No. 500643750 (MU 2686/15-1). The International Audio Laboratories Erlangen are a joint institution of the Friedrich-Alexander-Universität Erlangen-Nürnberg (FAU) and Fraunhofer Institute for Integrated Circuits IIS.
% This work was funded by X under Grant No. xxxx.
% and Grant No. yyyy.

% This work was funded by the Deutsche Forschungsgemeinschaft (DFG, German Research Foundation) under Grant No. xxxx and Grant No. yyyy.
%
% The International Audio Laboratories Erlangen are a joint institution of the Friedrich-Alexander-Universität Erlangen-Nürnberg (FAU) and Fraunhofer Institute for Integrated Circuits IIS. 

\end{acknowledgments} 

%%%%%%%%%%%%%%%%%%%%%%%%%%%%%%%%%%%%%%%%%%%%%%%%%%%%%%%%%%%%%%%%%%%%%%%%%%%%%
%bibliography here
% \bibliography{referencesMusic, referencesNew}

\end{document}